\documentclass[12pt,preprint]{aastex}

\begin{document}

\title{Identifying high redshift AGNs using X-ray hardness}

\author{J. X. Wang\altaffilmark{1,3}, S. Malhotra\altaffilmark{2}, 
J. E. Rhoads\altaffilmark{2}, C. A. Norman\altaffilmark{1,2}}
\begin{abstract}
The X-ray color (hardness ratio) of optically undetected X-ray sources
can be used to distinguish obscured active galactic nuclei (AGNs) at
low and intermediate redshift from viable high-redshift (i.e., $z > 5$) AGN
candidates.  This will help determine the space density, ionizing
photon production, and X-ray background contribution of the earliest
detectable AGNs.
High redshift AGNs should appear soft in X-rays, with hardness 
ratio HR $\sim$ -0.5, even if there is strong absorption by a 
hydrogen column density $N_H$ up to 
$10^{23} cm^{-2}$, simply because the absorption redshifts out of the soft 
X-ray band in the observed frame. Here the X-ray hardness ratio is defined 
as HR= (H-S)/(H+S), where S and H are the soft and hard band net counts 
detected by $Chandra$. High redshift AGNs that are Compton thick 
($N_H \gtrsim 10^{24} cm^{-2}$) could have HR $\sim$ 0.0 at $z >$ 5. 
However, these should be rare in deep Chandra images, since they have 
to be $\gtrsim$ 10 times brighter intrinsically, which implies 
$\gtrsim$ 100 times drop in their space density. 
Applying the hardness criterion (HR $<$ 0.0) can filter out 
about 50\% of the candidate high redshift AGNs selected from deep
Chandra images.
\end{abstract}


\altaffiltext{1}{Department of Physics and Astronomy, Johns Hopkins University, 
3400 N. Charles Street, Baltimore, MD 21218; jxw@pha.jhu.edu, norman@stsci.edu.}
\altaffiltext{2}{Space Telescope Science Institute, 3700 San Martin Drive, 
Baltimore, MD 21218; san@stsci.edu, rhoads@stsci.edu.} 
\altaffiltext{3}{Center for Astrophysics, University of Science and Technology of China, Hefei, Anhui 230026, P. R. China; jxw@ustc.edu.cn.}
\section {Introduction}

In the past few years, many deep $Chandra$ images of the extragalactic sky
have been obtained, with the 2 Ms $Chandra$ Deep Field North (CDF-N, e.g., 
Alexander et all. 2003) and 1Ms $Chandra$ Deep Field South (CDF-S, Giacconi 
et al. 2002; Rosati et al. 2002) being the two deepest.
Combining such exposures with deep optical images allows easy selection of 
candidate high redshift (i.e., $z > 5$, hereafter high-z) AGNs. 
Such high-z AGNs are extremely faint in optical bands blueward of
the Ly$\alpha$ wavelength, because of the
heavy absorption of UV light by the high redshift IGM (e.g., Fan et al. 
2001).
Optically undetected X-ray sources are thus good candidates for high-redshift
 AGNs. Their space density provides an upper limit on the density of high-z
AGNs, and can help determine their cosmological evolution and 
contribution to reionization (e.g., see Alexander et al. 2001; 
Barger et al. 2003a; Koekemoer et al. 2004; Wang et al. 2004). 
However, the absence of these candidates in optical bands makes 
them difficult to identify spectroscopically.
This motivates different approaches to studying them.
Currently, the efforts mainly focus on their infrared
colors (e.g., Yan et al. 2003; Koekemoer et al. 2004).

In this letter, we point out for the first time that the X-ray hardness 
ratio can be used to filter out low-z sources from these X-ray 
selected, optically undetected high-z candidates. High-z AGNs cannot be hard
in $Chandra$ images, because the absorption that makes the X-ray spectra harder 
is redshifted out of the soft X-ray band in the observed frame.
In section 2, we present detailed simulations to quantify this effect.
Here the hardness ratio (HR) is defined as (H-S)/(H+S), where S and H are
the soft (0.5 -- 2.0 keV) and hard (2.0 -- 8.0 keV)\footnote{In some papers, only the 2.0 -- 7.0 keV band
net counts were given (e.g., Giacconi et al. 2002; Stern et al. 2002; Wang
et al. 2004). The difference of the hardness ratios (HR) using
different hard bands (2.0 -- 8.0 keV or 2.0 -- 7.0 keV) is negligible
($\Delta$HR $<$ 0.008 from our simulations). This is actually expected because $Chandra$ has
much lower effective area above 7 keV and 7.0 -- 8.0 keV X-ray net count makes
a very small contribution to the whole hard band.
} X-ray band net counts detected by $Chandra$. 

\section{Simulations}

The X-ray spectra of low and intermediate redshift AGNs have been well
studied using the observations from several generations of X-ray satellites,
including EINSTEIN, ROSAT, ASCA, BeppoSAX, $Chandra$, and XMM. 
For type 1 AGNs (i.e., Seyfert 1 galaxies, and QSOs), the basic 
component of their X-ray spectra is a power law
with photon index $\Gamma \sim 1.9$ (e.g., Nandra et al. 1997; George
et al. 2000, Malizia et al. 2003) and an exponential cut off at high 
energies ($\sim$ 200 keV, see Malizia et al. 2003).
For type 2 AGNs (i.e., Seyfert 2 galaxies, and type 2 QSOs),
the power law is cut off at low energies
by photo-electric absorption, and the cutoff energy increases
with the column density of the intercepted torus (e.g., 
Turner et al. 1997a; Norman et al. 2002).
Recent $Chandra$ observations show that the X-ray spectra of QSOs
(i.e., luminous AGNs) at $z$ = 4.0 -- 6.3 are also 
well fitted by a power law with photon index $\Gamma$ = 1.9 (Vignali et al. 2004).
This indicates that although the space density of AGNs varies significantly
from $z \sim$ 0 -- 6, the shape of their intrinsic X-ray spectra evolves 
relatively little.
Evidence for warm absorbers and/or
soft excess emission (e.g., Krolik \& Kriss 2001; Piro, Matt \& Ricci 1997) have 
also been found in significant numbers of AGNs.
However, at high-z, these features shift out
of $Chandra$'s soft band.

In this section we present simulations to predict the X-ray colors
in $Chandra$ images for high-z AGNs by assuming a power law spectrum
($\Gamma$ = 1.9) with different absorption column densities 
(N$_H$ = 10$^{21}$, 10$^{22}$, 10$^{23}$, 10$^{24}$ cm$^{-2}$ respectively;
see Fig. 1 for the model spectra).
We used 
XSPEC 11.0.1
to do the simulations, and
model $wabs$ in XSPEC, a photoelectric absorption using Wisconsin 
cross-sections
(Morrison and McCammon, 1983), to simulate 
the neutral absorption in the rest frame. The 
$Chandra$ ACIS on-axis instrument response for CDF-S (Giacconi et al. 2002) 
was used, 
and the Galactic HI column density (N$_H$ = 0.8 $\times 10^{20} cm^{-2}$)
in CDF-S was taken account during the simulations\footnote{
Using slightly different Galactic HI column density, or $Chandra$ on-axis instrument response 
calculated for other ACIS-I fields, does not affect any results
presented in this paper.}. 
The output X-ray hardness ratios are plotted in Fig. \ref{hr}.
We can see that the predicted HR is a constant (-0.58 for 
$\Gamma$ = 1.9) for AGNs without intrinsic absorption at any
redshift, because of the power law shape of the
X-ray spectrum. The corresponding HR for 
different photon indices are also shown in Fig. \ref{hr}.
While the photon index for QSOs varies
from 1.5 to 3.0 (e.g., George et al. 2000),  the dominant source of variation
in hardness ratio HR is absorption. We show in  Fig. \ref{hr} 
that an extreme power-law 
 with $\Gamma  = 1.5$ would still give a soft color
(HR = -0.41).
For AGNs with intrinsic absorption, the predicted hardness ratio 
varies with redshift: at lower redshift, the X-ray spectra
are much harder because the soft X-ray emission are significantly attenuated
by the absorber; but at $z > 5$, 
we barely see differences between the X-ray hardness ratios of X-ray
spectra with absorption up to N$_H$ = 10$^{23}$ cm$^{-2}$, because 
the absorption has largely redshifted out 
of the soft X-ray band.
If the absorber is Compton thick (N$_H \ge 10^{24} cm^{-2}$), even the hard
X-ray emission would be significantly attenuated.
At $z > 5$, the predicted hardness ratio is $\sim$ 0.0.
In the compton thick regime (N$_H$ = $10^{24} cm^{-2}$), further correction to
photoelectric absorption is needed, due to  
Compton scattering (see Matt, Pompilio \& La Franca 1999; Yaqoob 1997).
Based on  figure 3 of Matt et al. (1999)
which includes compton scattering,
we conclude that the {\it shape}  of the transmitted curve is unchanged
by compton scattering, while the amplitude decreases
by a factor of 1.7.

For higher column density   (N$_H > 10^{24} cm^{-2}$), the direct X-ray
emission is strongly attenuated,
and the  X-ray spectra are dominated by a reflection component
from cold and neutral gas (e.g., Turner et al. 1997b). 
We used the XSPEC model $pexrav$ (Magdziarz \& Zdziarski 
1995) to simulate such pure reflection spectra.\footnote{The model of
Magdziarz \& Zdziarski is angle-dependent. The results we present in this
paper were derived by setting cos$\theta$ at 0.45, which is closest
in overall shape to the reflected spectrum averaged over all viewing angles.
Reflection from smaller viewing angles tends to be slightly harder ($\Delta HR < 0.1$), but 
obviously, this is not the case for type II AGNs which 
are supposed to edge-on.}
The Fe K emission line at 6.4 keV has a higher equivalent width (EW) in 
the reflection dominated X-ray spectra of AGNs (e.g., Ghisellini, Haardt \& 
Matt 1994; Levenson et al. 2002) since the direct component is absent.
Therefore we add
an Fe K emission line at 6.4 keV with EW of 1 keV in the rest frame,
which is normal in the
reflection dominated X-ray spectra of AGNs (e.g., Levenson et al. 2002). 

\section{Discussion}
The X-ray spectra of AGNs at  low to intermediate redshifts can be extremely 
hard due to heavy absorption ( with hardness ratios up to HR $\sim$ 1.0).  However, they are
much softer at higher redshift because the absorbed energy shifts out of the observed bands  (see Fig. 2 for 
simulations, and Fig. 12 of Szokoly et al. 2004 for the HR distribution
of a large sample of AGNs from $z$ = 0-4).
AGNs at $z \gtrsim 5$ with intrinsic absorption up to N$_H = 10^{23} cm^{-2}$ 
should have HR $\sim$ - 0.5, and the Compton-thick ones (N$_H \gtrsim 10^{24}
cm^{-2}$) should have HR $\lesssim$ 0.1.
Due to  heavy absorption, and compton scattering, the X-ray flux of 
Compton-thick AGNs (N$_H \sim 10^{24}$ $cm^{-2}$) is attenuated by a factor of 9.5
(See Fig. 1).

For pure reflection spectra, the attenuation is even larger:
assuming a reflection efficiency of 3\% in rest frame 2.0 -- 10.0 keV band
(e.g., see Norman et al. 2002) yields a factor of 21.
Thus any high redshift sources detected with a large hardness ratio would have intrinsic
X-ray luminosity of $\sim 10^{45}$ erg s$^{-1}$.
We know that brighter QSOs are much rarer; according to the X-ray luminosity 
function of AGNs (e.g., see Miyaji, Hasinger, \& Schmidt 2001; Ueda et al. 2003), 
a tenfold increase in luminosity
implies a hundredfold drop in the space density.
Furthermore, there is evidence that the fraction of type 2 AGNs 
decreases at higher intrinsic 
luminosity (e.g., Steffen et al. 2003; Ueda et al. 2003).
We conclude that those candidates with $HR \gtrsim 0.0$ are statistically unlikely to
be at $z \gtrsim 5$. They are either obscured AGNs or QSOs at low to
intermediate redshift.

$Chandra$ has detected a number of AGNs at high redshift.
Presently, there are 66 AGNs\footnote{See the excellent Web site 
http://wwww.astro.psu.edu/users/niel/papers/highz-xray-detected.dat 
maintained by Niel Brandt and Christian Vignali for the list of the 
high-redshift AGN detected in X-rays so far} at $z > 4$ detected 
by $Chandra$ ACIS, and 41 of them have published
soft (0.5 -- 2.0 keV) and hard (2.0 -- 8.0 keV) band net counts (or 0.5 -- 
2.0 keV band and 0.5 -- 8.0 keV band net counts)
(Alexander et al. 2003; Barger et al. 2002; Brandt et al. 2001, 2002;
Castander et al. 2003; Vignali et al. 2001, 2003a, 2003b; Bassett et al. 2004).
All of the 41 sources have HR $\le$ 0.0, with an average value of -0.60$\pm0.21$,
in excellent agreement with our estimates above.
All of these sources are type 1
AGNs, and most of them are optically selected. 
Since strong X-ray emission is expected from both type 1 and type 2 AGNs,
we expect the X-ray selected high-z AGN sample to include both
types.
However we argue that the
hardness ratio distribution of the X-ray selected high-z AGNs should be similar
to that of the known $z>4$ AGNs based on following reasons: I) the 3 X-ray selected AGNs with $z > 4$ have
consistent soft X-ray colors with the rest; II) high-z type 2 AGNs with N$_H$ 
upto 10$^{23} cm^{-2}$ are also expected to be X-ray soft with HR $\sim$ -0.5; 
III) type 2 AGNs with heavier absorption are much fainter in observed X-ray 
fluxes, thus are much rarer.

Wang et al. (2004) presented 168 X-ray sources detected by an 172 ks
$Chandra$ ACIS exposure in the Large Area Lyman Alpha (LALA, e.g,
Rhoads et al. 2003) Bo\"{o}tes field. 19 of them are not detected
in deep R and bluer band images ($R > 25.7$, Vega mag) 
and are possible $z \gtrsim 5$ objects.
The sources and their hardness ratios are listed in Table 1 of Wang et al.
In Fig. 3, we plot their hardness
ratios comparing with the 41 $Chandra$ detected $z > 4$ AGNs.
The two distributions are distinguishable at 99.99\% level
according to the Kolmogorov-Smirnov (K-S) test. 
Down to a flux limit of 1.7 $\times$ 10$^{-15}$ ergs cm$^{-2}$ s$^{-1}$,
the 19 sources
contribute $\sim$ 8\% to the total 2 -- 10 keV band X-ray background
(Wang et al. 2004).
After removing sources with HR $>$ 0, we lose 50\% of the candidates
and the contribution to the total X-ray background drops to $\sim$ 4\%.

Yan et al. (2003) studied the infrared colors of 6 $R$ band nondetected X-ray 
sources in CDF-S. 
These sources are listed in table 1. Two of them have HR $\gtrsim$ 
0.0, and cannot be at $z > 5$. This agrees with 
Yan's statement, that all these sources are unlikely to be at $z > 5$ based
on their infrared colors.
Koekemoer et al. (2004) presented 7 X-ray sources in CDF-S which are
undetected in deep multi-band GOODS HST ACS images, with extremely
high X-ray-to-optical ratios and red colors ($z_{850}$ -- K). 
These sources might
be located at $z >$ 6 such that even their Ly$\alpha$ emission is redshifted
out of the bandpass of ACS $z_{850}$ filter.
We find that 3 of the 7 sources have HR $\gtrsim$ 0.0,
indicating that their X-ray colors are too hard to be at $z > 6$.
These 3 sources could instead be type 2 AGNs at low to intermediate redshift.
Their nuclear optical emission should be heavily obscured, 
and their host galaxies need to be substantially underluminous, or 
dust-obscured, compared to other known sources (Koekemoer et al. 2004).
This confirms that there is a population of AGNs at low to
intermediate redshift which are extremely red, with high X-ray-to-optical
ratios, and undetected at the depth of GOODS.
The analogous sample among non-active galaxies is the population of
extremely red objects (EROs), which have surface density comparable to
Lyman break galaxies but a much lower typical redshift (e.g., Vaisanen
\& Johansson 2004).

Using deep multicolor optical data, Barger et al. (2003a) searched 
candidate $z > 5$ AGNs in the 2 Ms X-ray exposure of the CDF-N, and
found that besides the one X-ray source spectroscopically confirmed
at $z = 5.19$, only 31 X-ray sources with $z' >$ 25.2 and no
B or V band detection could lie at $z > 5$. Barger et al. (2003b) provided
multiband photometry for the CDF-N X-ray sources, which allows us
to identify the 31 candidate high-z AGNs\footnote
{Because the magnitudes discussed in Barger et al. 2003a and 2003b were measured 
with different aperture diameters, it is difficult to pick up exactly the same
31 sources discussed in Barger et al. 2003a based on the magnitudes provided
in Barger et al. 2003b. 
We identify 31 candidate high-z AGNs in CDF-N following the same manner
discussed in Barger et al. 2003a (i.e., z' band faint and $B,V$ band 
undetected), and tune the threshold magnitudes to match the numbers
of sources discussed in Barger et al. 2003a. Thus the sample we picked up
could statistically represent that of Barger et al. 2003a}. The hardness ratio distribution
of the 31 sources is plotted in Fig. 3.  15 of the 31 sources have hardness ratios 
HR $>$ 0.0, and thus cannot have  $z > 5$.
This directly supports the deduction that the majority of the optically
undetected X-ray sources are extreme examples of 
the optically faint X-ray source population, most of which are obscured AGNs
at z $\le$ 3 (Alexander et al. 2001).
Barger et al. (2003a) pointed out that Haiman \& Loeb (1999) overestimated
the surface density of $z > 5$ AGNs by at least an order of magnitude, and
similar conclusions can be seen in Alexander et al. (2001) and
Szokoly et al. (2004). Our analyses indicate that applying the X-ray hardness 
ratio cutoff (HR $>$ 0.0) could further reduce the surface density of candidate
$z > 5$ AGNs, and strengthen the above conclusions by a factor of 2.
This also supports the statement that AGNs made little contribution to 
the reionization  at $z \sim 6$ (Barger et al. 2003a; also see 
Dijkstra, Haiman \& Loeb 2004; Moustakas \& Immler 2004).

\section {Conclusions}

In this letter we present detailed simulations showing that high-z AGNs
cannot be hard in $Chandra$ images since X-ray absorption will shift out
of soft band at high redshift.
High redshift AGNs should appear soft in X-rays, with hardness
ratio HR $\sim$ -0.5 at $z \gtrsim 5$, even if there is strong absorption 
with $N_H$ up to $10^{23} cm^{-2}$. 
High-z AGNs that are Compton thick ($N_H \gtrsim 1
0^{24} cm^{-2}$) could have HR $\sim$ 0.0.  However, these should be rare 
in deep Chandra images, since they have to be $\gtrsim$ 10 times brighter 
intrinsically, which implies $\gtrsim$ 100 times drop in their space density. 
Most optically undetected X-ray sources with HR $\gtrsim$ 0.0 should be
obscured AGNs at low to intermediate redshift.
Applying the hardness criterion (HR $<$ 0.0) can filter out about 50\% of the 
candidate high redshift AGNs selected from deep Chandra images. 
This criterion can thereby help us to understand the
nature of these $Chandra$ X-ray sources, put additional robust constraints
to the space density of high-z AGNs, and significantly reduce the
expensive telescope time needed to spectroscopically confirm
high-z AGN samples based on deep $Chandra$ images.

\acknowledgements 
We would like to thank Dr. A. Hornschemeier and T. Yaqoob for helpful
discussions. The work of JW was supported by the CXC grant GO2-3152x and GO3-4148X.
We also would like to thank the referee for a prompt and helpful report.

\begin{deluxetable}{lccc}
\tablecaption{CDF-S X-ray sources in Koekemoer et al. 2004 and Yan et al. 2003}
\tablecolumns{9}
\tablewidth{0pt}
\tablehead
{
\colhead {ID$^a$} & \colhead {R.A(J2000)} & \colhead {Dec.(J2000)} & \colhead{HR$^b$}
}
\startdata
Koekemoer et al. 2004\\
66&   3:32:08.39&-27:40:47.0& $ 0.30^{+0.17}_{-0.17}$\\
69$^c$&   3:32:08.89&-27:44:24.3& $-0.71^{+0.52}_{-0.29}$\\
93&   3:32:13.92&-27:50:00.7& $-0.32^{+0.11}_{-0.68}$\\
133&  3:32:20.36&-27:42:28.5& $-0.01^{+0.28}_{-0.30}$\\
161&  3:32:25.83&-27:51:20.3& $-0.29^{+0.21}_{-0.71}$\\
191&  3:32:33.14&-27:52:05.9& $-0.35^{+0.11}_{-0.65}$\\
216&  3:32:51.64&-27:52:12.8& $0.15^{+0.22}_{-0.23}$\\
Yan et al. 2003\\
98$^d$&   3:32:14.67&-27:44:03.4& 0.09\\
140&  3:32:22.44&-27:45:43.9& $-0.39^{+0.25}_{-0.61}$\\
188&  3:32:32.17&-27:46:51.4& $0.39^{+0.15}_{-0.16}$\\
214&  3:32:38.03&-27:46:26.2& $-0.45^{+0.08}_{-0.08}$\\
222&  3:32:39.06&-27:44:39.1& $-0.11^{+0.10}_{-0.10}$
\enddata
\tablenotetext{a}{X-ray Source ID in Giacconi et al. (2002).}
\tablenotetext{b}{
The X-ray net counts in different bands used to calculate HR are from 
Alexander et al. (2003). We subtract the soft band net counts from the
total band (0.5 -- 8.0 keV) net counts to calculate the hard band net 
counts for 
sources with only upper limits of the hard band net counts in Alexander et al.
Errors for this quantity are calculated following the "numerical method"
described in \S1.7.3 of Lyons 1991.
}
\tablenotetext{c}{Also included by Yan et al. 2003.}
\tablenotetext{d}{Only 0.5 -- 8.0 keV band net count is available in Alexander
et al., and the 
hardness ratio is derived using the upper limit of the soft and hard band 
counts, the sum of which is very close to the total count (32.2 vs 27.9).
}
\end{deluxetable}

\begin{figure}
\plotone{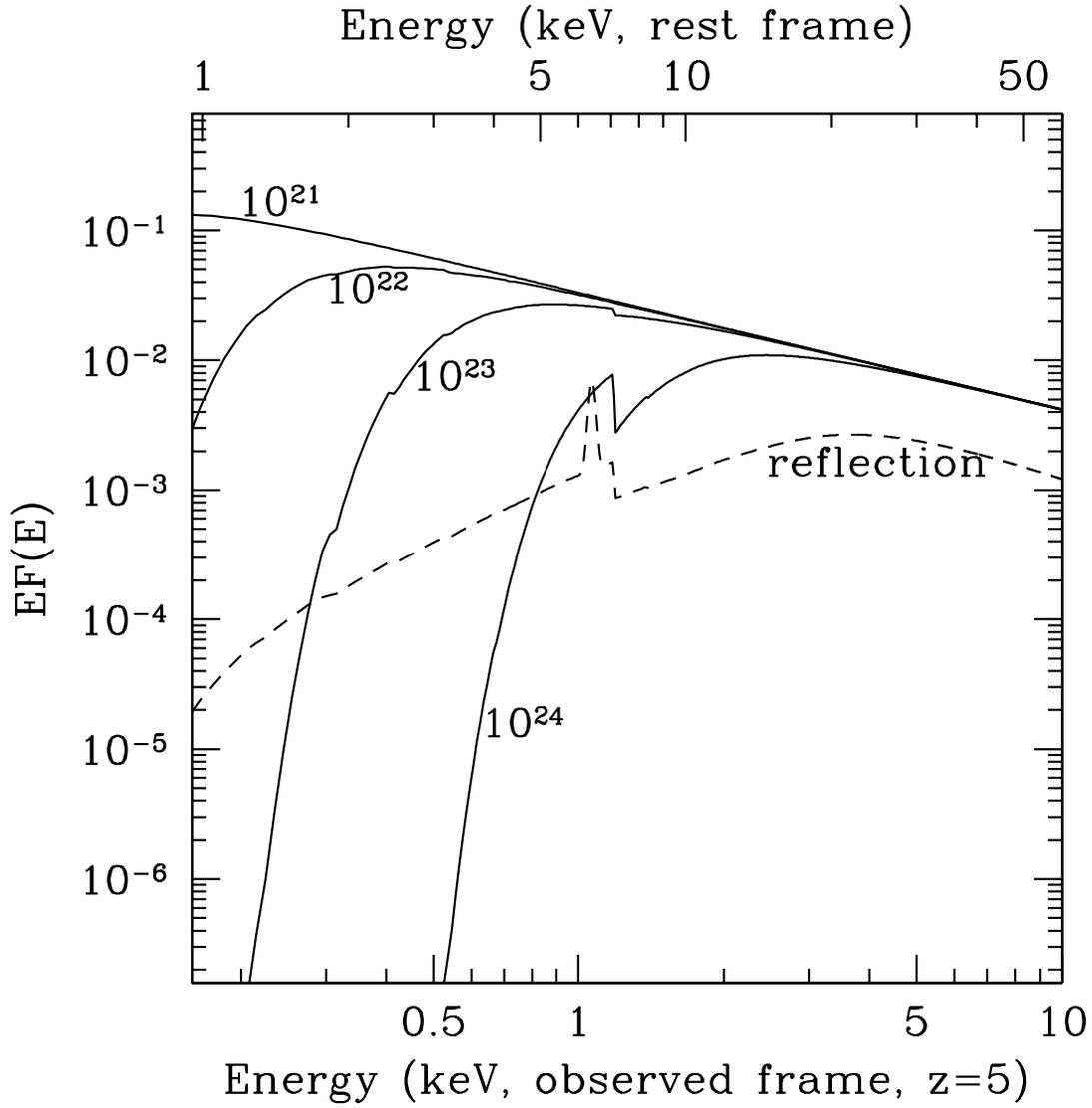}
\caption{
The input model spectra for N$_H$ = 10$^{21}$, 10$^{22}$, 10$^{23}$, 10$^{24}$ cm$^{-2}$,
and pure reflection spectrum. 
}
\label{spectra}
\end{figure}

\begin{figure}
\plotone{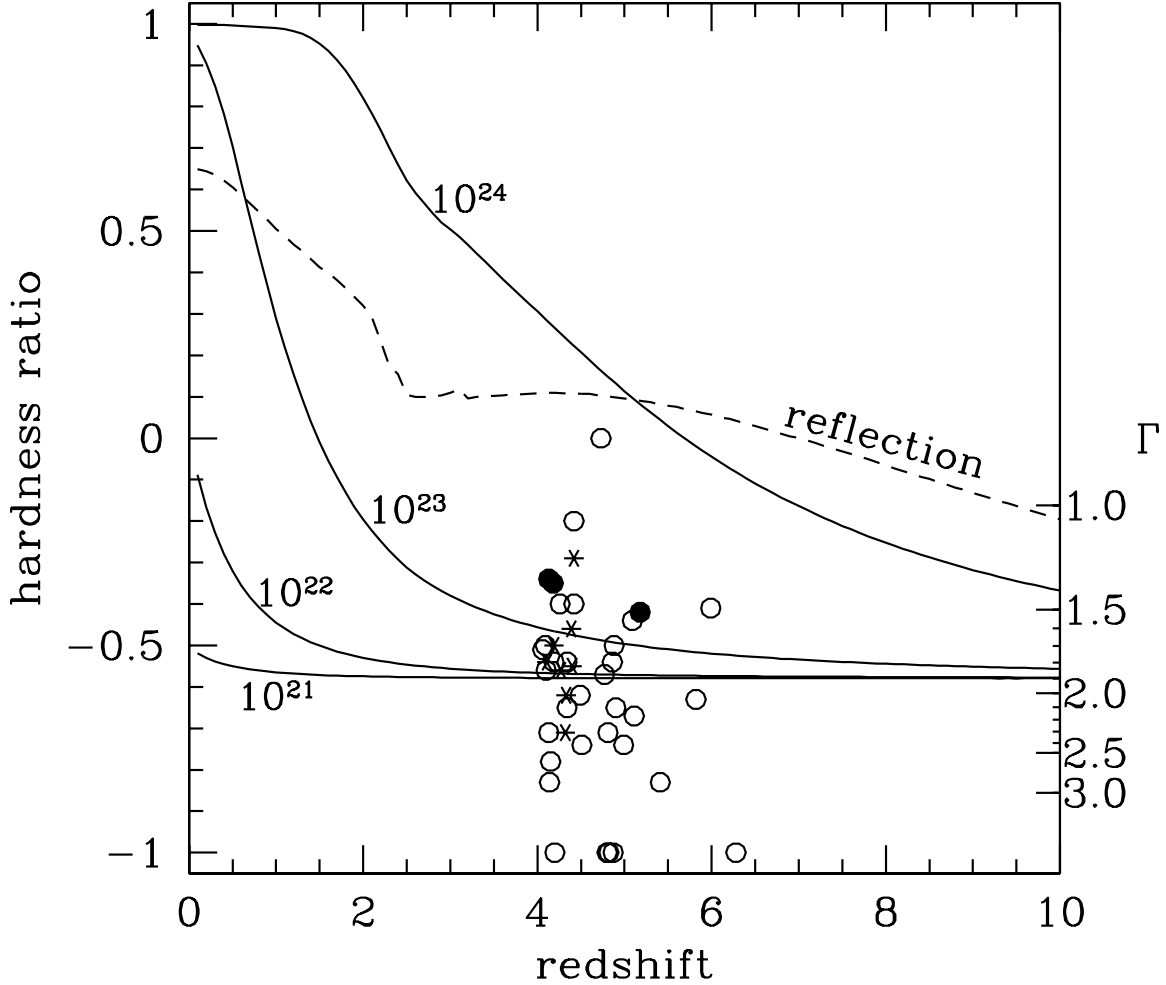}
\caption{
The predicted X-ray hardness ratio HR at different redshift.
The input spectrum is a power law with photon index
$\Gamma$ = 1.9, and absorbed by different 
column densities in the rest frame (N$_H$ = 10$^{21}$, 10$^{22}$, 10$^{23}$,
and 10$^{24}$ cm$^{-2}$ respectively) and pure reflection spectrum.
The output is the 
hardness ratio, if observed by $Chandra$. The hardness ratio
HR is defined as (H-S)/(H+S), where H and S are $Chandra$ net
counts in the soft (0.5 -- 2.0 keV) and hard (2.0 -- 8.0 keV)
X-ray band. 
Along the right ordinate, we mark the photon indices $\Gamma$ of absorption-free
power-law spectra which could reproduce the corresponding
hardness ratios. 
The known AGNs at $z > 4$ detected by $Chandra$ are also marked: 
open circles are optically selected, solid
circles are X-ray selected, and stars are radio selected.
See Fig. 12 of Szokoly et al. 2004 for a sample of Chandra detected
AGN at $z < 4$.
}
\label{hr}
\end{figure}

\begin{figure}
\plotone{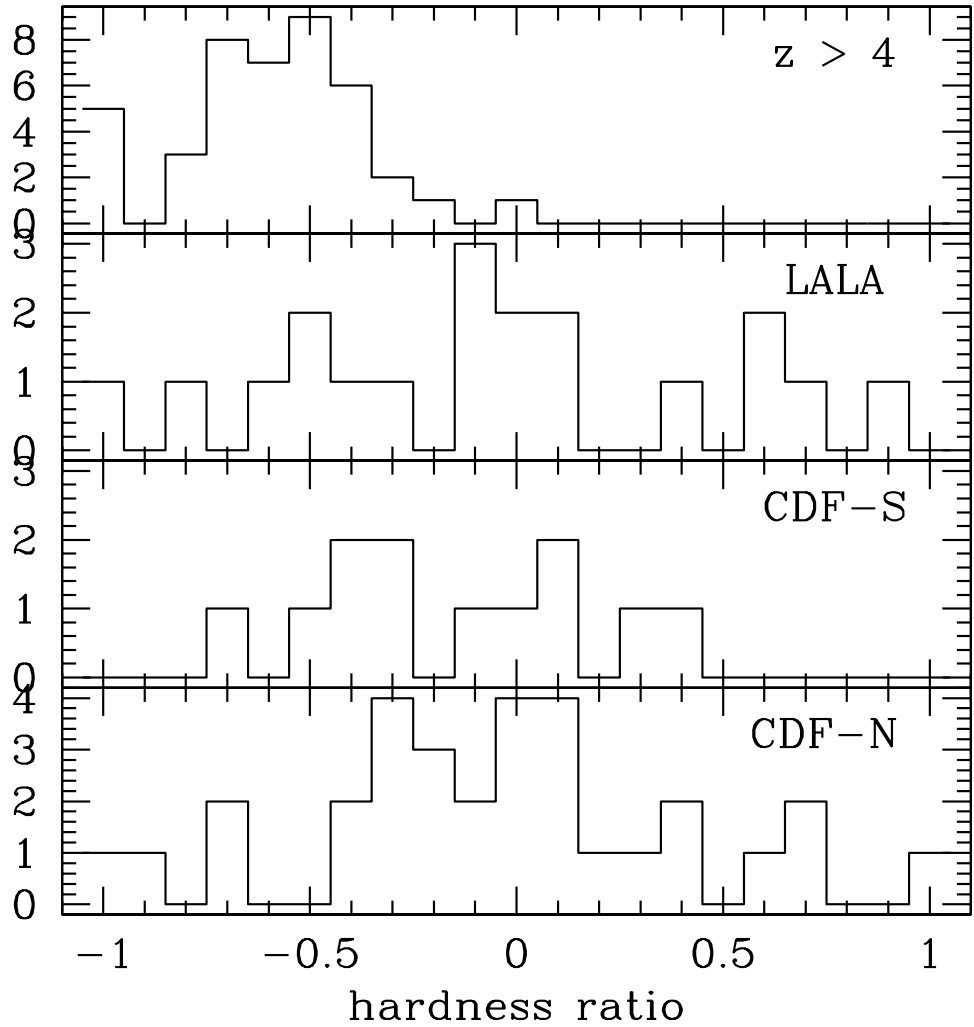}
\caption{
The X-ray hardness ratio distributions of 41 $Chandra$ detected 
$z > 4$ AGNs ($z>4$), 19 high-z candidates in LALA Bo\"{o}tes field
(LALA), 12 CDF-S sources in Koekemoer et al. (2004) and Yan et al (2003),
and 31 CDF-N sources discussed in Barger et al. (2003a). The later three
distributions are significantly different from the first one at the 
level $>$ 99.99\% based on the K-S test.
See the text for details.}
\end{figure}

\end{document}